\documentclass[%
 reprint,
 nofootinbib,
 amsmath,amssymb,
 aps,
]{revtex4-2}

\usepackage{graphicx}
\usepackage{dcolumn}
\usepackage{bm}
\usepackage{xcolor}
\usepackage{amsmath}
\usepackage{mathrsfs}
\usepackage{amssymb}

\begin{document}

\title{Entering the overcritical regime of nonlinear Breit-Wheeler pair production
in collisions of bremsstrahlung $\gamma$-rays and superintense, tightly focused laser pulses}

\author{I. \surname{Elsner}}
\email{ingo.elsner@hhu.de}
\affiliation{Institut f\"{u}r Theoretische Physik I, Heinrich-Heine-Universit\"{a}t D\"{u}sseldorf, Universit\"{a}tsstra\ss{e} 1, 40225 D\"{u}sseldorf, Germany.}
\author{A. \surname{Golub}}
\email{alina.golub@hhu.de }
\affiliation{Institut f\"{u}r Theoretische Physik I, Heinrich-Heine-Universit\"{a}t D\"{u}sseldorf, Universit\"{a}tsstra\ss{e} 1, 40225 D\"{u}sseldorf, Germany.}
\author{S. \surname{Villalba-Ch\'avez}}
\email{selym@tp1.hhu.de}
\affiliation{Institut f\"{u}r Theoretische Physik I, Heinrich-Heine-Universit\"{a}t D\"{u}sseldorf, Universit\"{a}tsstra\ss{e} 1, 40225 D\"{u}sseldorf, Germany.}
\author{C. \surname{M\"{u}ller}}
\email{c.mueller@tp1.hhu.de}
\affiliation{Institut f\"{u}r Theoretische Physik I, Heinrich-Heine-Universit\"{a}t D\"{u}sseldorf, Universit\"{a}tsstra\ss{e} 1, 40225 D\"{u}sseldorf, Germany.}

\date{\today}

\begin{abstract}
Near-future high-intensity lasers offer prospects for the observation of nonlinear Breit-Wheeler pair production in an overcritical field regime, where the quantum nonlinearity parameter substantially exceeds unity. This experimentally yet unexplored scenario is envisaged here to be reached via the collision of a tightly focused laser pulse with high-energy bremsstrahlung photons. We calculate the achievable number of pairs in a range of laser intensities around 10$^{23}$ W/cm$^2$ and GeV-energies of the incident bremsstrahlung-generating electron beam. We investigate under which conditions the attenuation of the $\gamma$-beam due to the production process must be taken into account and how much the second generation of created pairs contributes to the total yield. In the considered interaction regime, where the local production rate grows rather moderately with higher field intensities, it is shown that the range of mostly contributing bremsstrahlung frequencies is generally very broad. For sufficiently large values of the quantum nonlinearity parameter, an optimum domain of frequencies emerges which is located far below the spectral endpoint. Furthermore, we show that it is beneficial for achieving the optimum pair yield to increase the interaction volume by a wider laser focus at the expense of decreased field intensity.
\end{abstract}

\maketitle


\section{Introduction}

Creation of electron-positron pairs from vacuum in the presence of very strong electromagnetic fields belongs to the most dramatic predictions of quantum electrodynamics. After a few seminal studies in the early days of relativistic quantum mechanics \cite{Sauter}, the availability of laser radiation starting in the 1960s has sparked sustained interest of theoreticians in the subject \cite{NikishovRitus1,NikishovRitus2,Reiss1962, Reiss1971, Popov, Mostepanenko, RitusReview}.

In particular, a nonlinear version of the Breit-Wheeler effect was studied (see \cite{RitusReview, DiPiazzaReview, FedotovReview} and references therein) where pairs are produced by multiphoton absorption in the collision of a high-energy $\gamma$-photon with an intense laser wave, according to the reaction $\omega^\prime + n\omega \to e^+e^-$, where $n$ denotes the number of absorbed low-frequency laser photons. Various interaction regimes of the process have been identified that are distinguished by the values of two dimensionless parameters: on the one hand, the classical intensity parameter $\xi=\vert e\mathcal{E}_0\vert/(m\omega)$, with the laser frequency $\omega$ and its peak field strength $\mathcal{E}_0$; and on the other hand, the quantum nonlinearity parameter given by $\kappa=2\omega^\prime \mathcal{E}_0/(m E_c)$, with the critical Schwinger field $E_c=m^2/\vert e\vert\approx 1.3 \times 10^{16}$~V/cm and assuming a counterpropagating beam geometry. Besides, $e$ and $m$ denote the electron charge and mass, respectively. The theoretical treatment in the seminal papers \cite{NikishovRitus1,NikishovRitus2,Reiss1962} relied on monoenergetic $\gamma$-photons and a plane-wave description of the laser field. In the 1990s, the nonlinear Breit-Wheeler process has been observed experimentally in a few-photon regime ($n\sim 5$) at $\xi\lesssim 1$, where the process rate was found to follow a power law of the form $R \sim \xi^{2n}$ \cite{Burke}.

Due to the ongoing progress in high-intensity laser technology, the experimental interest in the nonlinear Breit-Wheeler process is currently revived \cite{E320, LUXE, CALA, ELI} -- which is accompanied by corresponding theoretical efforts \cite{Heinzl2010, Kaminski/Krajewska, Titov2012, Selym, Meuren/DiPiazza, Jansen-Spin, Grobe2018, Titov2020, Lu2020, Seipt2020, Heinzl2020, Wan2020, Podszus, Chen2023, Mahlin2023}. The aim is to probe the hitherto unobserved regimes with $\xi \gg 1$ where the rate exhibits a manifestly nonperturbative dependence on the external field strength.  For $\kappa\ll 1$, it has a tunneling-like exponential form $R\sim \mathrm{e}^{-8/(3\kappa)}$, whereas for $\kappa \gg 1$ the rate scales as $R\sim \kappa^{2/3}$ \cite{RitusReview}. Since  strong focusing is required to achieve laser pulses with such high intensities,  it is crucial in calculations to properly account for the spatiotemporal structure of the fields \cite{DiPiazzaFocus, Riconda, Golub2022, Eckey2024}.

A promising possibility to generate high-energy $\gamma$-rays is offered by bremsstrahlung of highly relativistic electrons from high-$Z$ targets, as proposed by Reiss in the 1970s \cite{Reiss1971}, and recently further investigated by several groups \cite{Blackburn2018, Hartin, Riconda, Eckey2022, Golub2022, Eckey2024, Jeong2023, King2024}. In this situation, the $\gamma$-photons are not monoenergetic but span a broad range of energies. This route is followed in the experimental designs for nonlinear Breit-Wheeler pair creation at CALA \cite{CALA} and DESY \cite{LUXE}. If successful, they could be considered as 'discovery experiments' for the highly nonlinear regime of the strong field Breit-Wheeler process, which will be followed afterwards by 'precision experiments' that aim at covering large regions of the parameter space with enhanced statistical accuracy. A detailed theoretical analysis of the experiment planned at CALA has recently been carried out \cite{Golub2022, Eckey2024}, including laser focusing effects and the broad frequency spectrum of bremsstrahlung. This experiment is going to operate in the $\xi\gg 1$ regime with  $\kappa \sim 1$, exploiting laser wakefield accelerated incident electrons of about 2.5 GeV energy to generate the bremsstrahlung in a tungsten target \cite{CALA}.

In the present paper, we extend the previous analysis \cite{Golub2022} to the overcritical regime of large $\kappa$ values up to $\kappa\approx 30$. It could, for instance, be realized in the future at the Extreme Light Infrastructure (ELI) \cite{ELI} or CoReLS \cite{CoReLS} facilities. Our main goal is to reveal characteristic differences between the intermediate ($\kappa\approx 1$) and overcritical ($\kappa\gg 1$) regimes of the nonlinear strong-field  Breit-Wheeler process with focused laser pulses and $\gamma$-photons from bremsstrahlung.\footnote{The $\kappa$ values considered here remain far below the domain where perturbation theory with respect to the quantized radiation field is expected to break down, i.e. $\kappa\ll\alpha^{-3/2}\approx 1.6\times 10^3$, with $\alpha$ referring to the fine-structure constant \cite{RitusReview, DiPiazzaReview, FedotovReview, Ritus1972}.} We will show that the overcritical field regime requires an adjusted theoretical approach and exhibits qualitatively distinct properties in terms of its scaling behaviors. This affects, in particular, its dependencies on the laser pulse duration and focusing as well as the mainly contributing spectral range of $\gamma$-photons. Our outcomes can be relevant for the design of future experiments.

\begin{figure}[ht]
\includegraphics[width=0.5\textwidth]{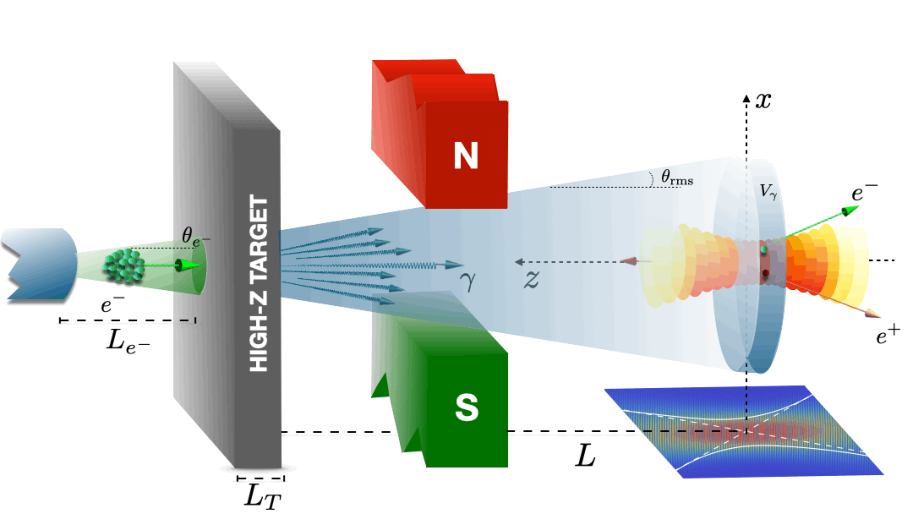}
\caption{\label{FigSelym} Scheme of the experimental setup for entering the overcritical regime of the nonlinear Breit-Wheeler pair production process from the collision of bremsstrahlung $\gamma$-photons and a high-intensity laser pulse. The collision of a monoenergetic electron beam with a narrow high-$Z$ target will result in the bremsstrahlung burst utilized in the setup. A magnet  after the target will deflect residual electrons away from the site where the photon-photon collision occurs.}
\end{figure}

It is worth noting that the overcritical regime of strong-field pair production can also be probed when an incident high-energy electron beam interacts directly with an ultrastrong laser pulse \cite{E320,Ilderton2019,Baumann2019}. In this case, however, the electron beam can be strongly distorted by radiation reaction effects before reaching the focal region of highest field strength. In this regard, the use of a target foil that first converts the incident electron energy in a controled way into $\gamma$-photons, which afterwards penetrate into the laser focal region, offers certain advantages. Figure~\ref{FigSelym} shows a scheme of the envisaged setup.

Our study  is presented as follows: In Sec.~\ref{Theory} we first briefly review the theoretical description of Breit-Wheeler pair creation in collisions of bremsstrahlung $\gamma$-rays with a focused high-intensity laser pulse from Ref.~\cite{Golub2022} and, afterwards, extend it to the overcritical regime. Next, in Sec.~\ref{Results}, our numerical results obtained for the overcritical regime $\kappa > 1$ are discussed and put in contrast to the properties of the pair creation process at $\kappa \approx 1$. Finally, the conclusions are drawn in the last section. Throughout the manuscript, the Lorentz-Heaviside units [$c=\hbar=\epsilon_0=1$] and the metric tensor with the signature $\mathrm{diag}(g^{\mu \nu})=(1,-1,-1,-1)$ are used.

\section{Theoretical treatment}\label{Theory}

As the discussion in this manuscript aims at pair production at high laser intensities with $\xi \gg 1$, the locally constant field approximation is applicable  \cite{RitusReview, Seipt2020}. Under such a condition, the formation length $l\sim \lambda/(\xi\pi)$ is much smaller than the laser wavelength $\lambda=2\pi\omega^{-1}$, evidencing that the pairs are created in regions where the background is nearly a constant crossed field. Accordingly, the rate of the process per $\gamma$-photon can be approximated locally by the one linked to pair production in a crossed field configuration \cite{NikishovRitus1,RitusReview} 
\begin{equation}
\label{R}
R(\kappa) = -\frac{\alpha m^2}{6\sqrt{\pi}\omega'} \int_{1} ^\infty \frac{du (8u +1)}{u\sqrt{u(u-1)}} \frac{\Phi'(z)}{z},
\end{equation}
where the local value $\kappa=\kappa(\vec{r},t)$ of the quantum nonlinearity parameter at the considered space-time point must be inserted. Here, $\alpha=e^2/(4\pi)\approx 1/137$ is the fine structure constant, $z=(4u/\kappa )^{2/3}$ and $\Phi'(z)$ stands for the derivative of the Airy function $\Phi(z) = \frac{1}{\sqrt{\pi}}\int_0^\infty dt \cos(\frac{t^3}{3}+zt)$.

Equation \eqref{R} forms the basis for two different theoretical approaches to strong-field Breit-Wheeler pair production that will be outlined in Secs.~\ref{TheorySubsectionB} and \ref{TheorySubsectionC} below. They are used to calculate the first generation of  pairs, which are created directly from the incident $\gamma$-photons. An extension of the approach, which allows us to include also the second generation of pairs from a shower-like process, will be described in Sec.~\ref{TheorySubsectionD}. Beforehand, we shall specify the colliding high-intensity focused laser pulse and high-energy bremsstrahlung $\gamma$-rays. 

\subsection{Descriptions of the focused laser pulse and bremsstrahlung spectrum} \label{TheorySubsectionA}
The laser field shall by taken as a tightly focused Gaussian pulse in the paraxial approximation. A geometry is chosen in which this strong background propagates in $z$-direction and is linearly polarized along the $x$-axis. The corresponding electric field reads \cite{Salamin}
\begin{equation}\label{fieldparaxial}
E_x = \mathcal{E}_0\frac{\mathrm{e}^{-\left(\sqrt{2\ln(2)}\frac{(t-z)}{\tau}\right)^2}}{\sqrt{1+\zeta(z)^2}}\, \mathrm{e}^{-\left(\frac{\rho}{w(z)}\right)^2}\mathrm{sin}(\Phi),
\end{equation} 
where $\tau$ stands for the pulse duration (at FWHM from the intensity), $\rho^2=x^2+y^2$, $\zeta=z/z_R$ with the Rayleigh length $z_R=\pi w_0^2/\lambda$ and the beam waist radius at the focus $w_0$, whereas $w(z)=w_0\sqrt{1+\zeta(z)^2}$. Moreover, $E_x=B_y$ holds and the field phase in Eq.~\eqref{fieldparaxial} reads
\begin{equation}\label{Phi}
\Phi=\omega(t-z)-\zeta(z)\frac{\rho^2}{w^2(z)}+\mathrm{arctan}(\zeta).
\end{equation} 
It is worth remarking that the spacetime dependent quantum nonlinearity parameter $\kappa = |e|\sqrt{-(F_{\mu\nu}k'^{\nu})^2}/m^3$, written in terms of the electromagnetic field tensor $F_{\mu\nu}$, turns out to be 
\begin{equation} \label{kappa}
\kappa(t,\rho,z) =  2\frac{|e|\omega'}{m^3}|E_x(t,\rho,z)|
\end{equation}
for a counterpropagating geometry, in which case, the interacting $\gamma$-quantum is characterized by a wave vector $k'=\omega'(1,0,0,-1)$.

In order to take into account bremsstrahlung, the corresponding  photon distribution has to be considered. For high energies $E_0$ of incident bremsstrahlung electrons---which is the case in the present paper as we assume GeV energies---the collimation angle of the photons emitted by a single electron can be approximated as an inverse Lorentz factor $\theta_{\gamma} \approx1/\gamma_e= m/E_0\sim \mathcal{O}(1)$ mrad. Hence, the vast majority of the produced bremsstrahlung photons is emitted in the direction of propagation of incident electrons. Consequently, the photons energy spectrum can be approximated by \cite{Tsai, PPG}
\begin{equation}\label{thin}
I_\gamma(f,\ell)\approx \frac{\ell}{f}\left(\frac{4}{3}-\frac{4f}{3}+f^2\right)
\end{equation} 
with the normalized photon energy $f = \omega'/E_0$ and the normalized target thickness $\ell = L_{\mathrm{T}}/L_{\mathrm{rad}}$, which relates the target thickness $ L_{\mathrm{T}}$ to the radiation length of the material $L_{\rm rad}$. Note that the formula above is valid within the complete screening approximation and for materials with high $Z$-numbers, as well as  for very thin targets with $\ell \ll 1$. Its divergence for $f\to 0$ is not harmful in the present context, because the low-energy part of the spectrum below $f \lesssim 0.01$ will practically not contribute to the pair creation in the considered parameter regime (see Sec.~\ref{Results}). Besides, its non-zero value at the spectral endpoint $f=1$ does not distort the calculated pair yields either, because the real bremsstrahlung spectrum would decline to zero very steeply at this point (see Ref.~\cite{Golub2022} for more details, in particular Fig.~6 therein).

\begin{figure}[ht]
\includegraphics[width=0.5\textwidth]{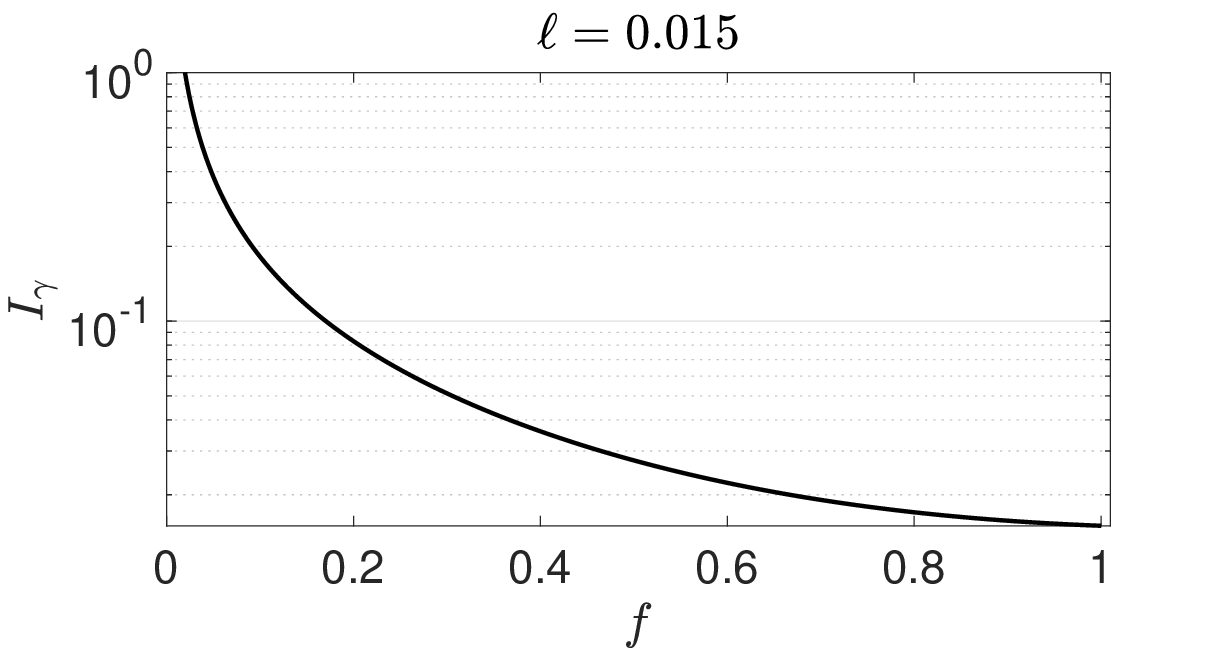}
\caption{\label{fig0} Bremsstrahlung spectrum as a function of the normalized photon energy $f=\omega'/E_0$ for a target with $\ell=0.015$.}
\end{figure}

Hereafter, a target made up of tungsten  is considered  with $L_{\rm rad}=3.5$ mm and $L_T = 50 \: \rm \mu m$, corresponding to $\ell = 0.015$ \cite{CALA}. In Fig.~\ref{fig0} one sees the bremsstrahlung spectrum per radiating incident electron as given in Eq.~(\ref{thin}) for the chosen target thickness as a function of $f$. We shall assume throughout that the incident electron bunch contains a total (absolute) charge of $Q_e = 10$ pC and that $\delta = 1$\% of the electrons will emit bremsstrahlung, which represents a reasonable fraction for the chosen target thickness \cite{CALA}. Accordingly, the total number of radiating electrons is $Q_e \delta/|e|  \approx 6\times 10^5$.

At this point, we can elucidate on the volume $V_{\gamma}$ and area $A_\gamma$, respectively, occupied by the $\gamma$-photons generated from bremsstrahlung. Taking into account that, in the considered experimental setup (see Fig.~\ref{FigSelym}), the incident electron beam spreads over a distance $L_{e^-}$ with the angle $\theta_{e^-}$ and that afterwards the generated beam of bremsstrahlung spreads over a distance $L$ to the interaction zone with the angle $\theta_{\mathrm{rms}}$, the volume of the $\gamma$-photons can be approximated by $V_{\gamma} \approx 2 z_R A_{\gamma}$, with the cross sectional area
\begin{equation}\label{vgamma}
A_{\gamma} \approx \pi (L_{e^-}\theta_{e^-}+L\theta_{\mathrm{rms}})^2,
\end{equation}
where $\theta_{\mathrm{rms}}=(\theta_{e^-}^2+\theta_{\gamma}^2)^{1/2}$ is the root-mean-square value of the electron-beam divergence and the angular spread of emitted bremsstrahlung for a single electron. Here, it is assumed that the electrons are accelerated via laser wakefield acceleration by utilizing a portion of the same laser seed from which the tightly focused laser pulse for the pair production is generated. Moreover, for the longitudinal extension of the electron beam a conservative value of $2z_R$ is supposed (i.e.~twice the Rayleigh length of the latter pulse). For further details, we refer to Sec.~\ref{TheorySubsectionA} in \cite{Golub2022}. In the further course of our consideration the electron beam spreading parameters are set to $L_{e^-} = 10$ cm and $\theta_{e^-}=0.5$~mrad (see \cite{CALA}). The distance to the interaction zone travelled by the $\gamma$-photons is taken as $L = 0.5$\,m. Accordingly, when the incident electron energy is $E_0 = 2.5$ GeV, the cross sectional area amounts to $A_\gamma \approx \pi (320 \mu{\rm m})^2$. It slightly shrinks to $A_\gamma \approx \pi (300\, \mu{\rm m})^2$ for $E_0 = 10$ GeV, because the beam of $\gamma$-photons is less divergent then.

\subsection{Pair production at moderate field strengths: Stable $\gamma$-beam model} \label{TheorySubsectionB}
In the following, we briefly describe the method of Ref.~\cite{Golub2022} (see especially Secs.~\ref{TheorySubsectionB} and \ref{TheorySubsectionC} therein) which allows us to calculate the number of created Breit-Wheeler pairs, when both the quantum nonlinearity parameter $\kappa$ and the laser pulse duration $\tau$ are not too large. We denote our corresponding approach as ``stable $\gamma$-beam model'' (S-model in short) for reasons that will become clear in Sec.~\ref{TheorySubsectionC}. 

In the considered situation, the number of pairs created by a $\gamma$-photon of energy $\omega'$ can be estimated directly by integrating the local rate in Eq.~\eqref{R} over the interaction time $T_{\mathrm{int}}$ and interaction volume  $V_{\mathrm{int}}$:
\begin{equation}\label{N}
\mathcal{N}(\omega') \approx \frac{1}{V_\gamma} \int_{V_{\mathrm{int}}} \int_{T_{\mathrm{int}}} R(\kappa)\vert_{\kappa \to \kappa (\vec{r},t)}.
\end{equation} 
Both domains $T_{\mathrm{int}}$ and $V_{\mathrm{int}}$ are supposed to be determined by the profile of the Gaussian laser pulse \eqref{fieldparaxial}. Since the latter has no dependence on the azimuthal angle and we assume a counterpropagating beam geometry, the number of created pairs can be written as 
\begin{equation}
\label{NGauss}
\mathcal{N}(\omega') \approx \frac{2\pi}{V_\gamma} \int_{-\infty}^{\infty}dt\int_0^{\rho_{\rm max}} \rho d\rho\int_{-t-z_R}^{-t+z_R}dz \ R(\kappa)\big\vert_{\kappa\to\kappa(t,\rho,z)}.
\end{equation} 
Since the transverse extension of the bremsstrahlung $\gamma$-beam is much larger than the transverse size of the laser pulse, the latter determines their overlap region and, thus, the upper boundary of the $\rho$ coordinate; in the calculations, the value $\rho_{\rm max}=3w_0$ has been taken. The boundaries of the $z$-integration take the spatial overlap between laser pulse and bremsstrahlung beam into account as well (see also Fig.~\ref{AlinaKappaPlot} below). In order to include the energy distribution of the bremsstrahlung, an incoherent average of Eq.~(\ref{NGauss}) over the spectrum \eqref{thin} must be taken. Thus, the number of created Breit-Wheeler pairs per radiating electron in our S-model is given by
\begin{equation}
\label{NGamma}
\begin{split}
N_S \approx  \int_0^1 df~\mathcal{N}(\omega')\,I_\gamma(f,\ell)\,.
\end{split}
\end{equation}
Before moving on to the next section we note that the substitution $z = - t - \tilde{z}$ transforms Eq.~\eqref{NGauss} into
\begin{equation}
\label{NGauss2}
\mathcal{N}(\omega') \approx \frac{2\pi}{V_\gamma} \int_{-\infty}^{\infty}dt\int_0^{\rho_{\rm max}} \rho d\rho\int_{-z_R}^{z_R}d\tilde{z} \ R(\kappa)
\end{equation}
with $\kappa = \kappa(t,\rho,-t-\tilde{z})$. In the transformed expression, the $\tilde{z}$-integration boundaries are decoupled from the time variable, in contrast to Eq.~\eqref{NGauss}. The performed setting $z = z(t) = - t - \tilde{z}$, which now links the longitudinal component with the time, offers a physical interpretation in semiclassical terms: it describes the (longitudinal coordinate of the) trajectory of a $\gamma$-photon moving with the speed of light in negative $z$-direction. The quantity $\tilde{z}$ correspondingly represents a relative longitudinal offset between different $\gamma$-photons. This offset runs between $-z_R$ and $z_R$, in accordance with the assumed extension of the bremsstrahlung $\gamma$-pulse. The picture of point-like $\gamma$-photons propagating along classical trajectories will be exploited further in the theoretical model of the next section.

\subsection{Pair production at very high field strengths: Decaying $\gamma$-beam model} \label{TheorySubsectionC}
The method described in Sec.~\ref{TheorySubsectionB} is only applicable when the quantum nonlinearity parameter $\kappa$ and the laser pulse duration $\tau$ are not too large. Otherwise this method will overestimate the pair production yield, because it disregards the attenuation of the $\gamma$-rays that arises from their decay into electron-positron pairs. In the following we present another calculational method that is able to describe the nonlinear Breit-Wheeler process at $\kappa$ values substantially exceeding unity. To this end, we adjust the treatment put forward in Ref.~\cite{Riconda} and extend it to incorporate $\gamma$-rays from bremsstrahlung with a broad frequency spectrum.

While in Sec.~\ref{TheorySubsectionB} the $\gamma$-photons were modelled from the beginning by a continuous number density $\varrho_\gamma$ that interacts with the full spatiotemporal extent of the laser pulse, the approach of the present section relies on the contributions of individual $\gamma$-photons to the pair production. We assume that a $\gamma$-photon moves with the speed of light ($c = 1$) along a straight-line trajectory
\begin{align} \label{traj}
\vec{r}(t)= -  t \hat{e}_z + b \hat{e}_\rho
\end{align}
where we have set the longitudinal coordinate $z(t)=-t$ antiparallely to the $z$-axis and introduced a transverse 'impact parameter' $b$ to the origin of the coordinate frame where the center of the laser pulse is located. The longitudinal offset $\tilde{z}=0$ has been set to zero [see Eq.~\eqref{NGauss2}]. This simplifying assumption, which facilitates the computations, shall be shown below to represent a good approximation in the considered regime of parameters. During its path through the laser field, a $\gamma$-photon of energy $\omega'$ decays into an electron-positron pair with the probability
\begin{align}
P(b,\omega') = 1- \text{exp} \left[- \int_{-\infty}^\infty dt\,R(\kappa)\right]_{\kappa\to \kappa[\vec{r}(t)]} \label{RicondaSinglePhotonProbability}
\end{align}
which coincides with the probability for producing a pair. In the exponent of Eq.~(\ref{RicondaSinglePhotonProbability}), the pair production rate $R(\kappa)$ enters, which is evaluated along the trajectory of the $\gamma$-photon. We note that $R(\kappa)$ is valid up to the first order in $\alpha$ [see Eq.~\eqref{R}]. Higher-order processes such as $\omega'+n\omega\to e^+e^- + \omega''$, where the pair production is accompanied by the emission of a $\gamma$-photon $\omega''$, are not included here. We will come back to this point in Sec.~\ref{TheorySubsectionD} below.

Clearly, when the time-integrated rate in Eq.~\eqref{RicondaSinglePhotonProbability} is much smaller than unity, the exponential can be expanded and one obtains
\begin{align}
P(b,\omega') \approx \int_{-\infty}^\infty dt\,R(\kappa) \big\vert_{\kappa\to\kappa[\vec{r}(t)]}
\label{Rt}
\end{align}
in accordance with the method of Sec.~\ref{TheorySubsectionA}. However, when the time-integrated rate is not small, which occurs for large values of $\kappa$ and/or $\tau$, the previous method overestimates the number of created pairs substantially. The reason is that a $\gamma$-photon which has decayed into a pair, is lost from the $\gamma$ beam and cannot induce further pair creation events. Accordingly, the approach of the present section shall be refered to as ``decaying $\gamma$-beam model'' (D-model). Thus, in a situation where Eq.~\eqref{RicondaSinglePhotonProbability} cannot be reduced to Eq.~\eqref{Rt}, the transition from the process rate to the process probability adopts a nonperturbative form.

The pair production probability $P(b,\omega')$ holds for a $\gamma$-photon of energy $\omega'$ passing the central point of the laser focus at distance $b$. In order to obtain the pair yield resulting from an extended beam of bremsstrahlung $\gamma$-photons, we first average the individual probability over the transversal interaction area according to
\begin{align}
\label{ricondaP}
\mathscr{P}(\omega') = \frac{2\pi}{A_\gamma} \int_0^{b_{\rm max}} db \: b\, P(b,\omega')\,.
\end{align}
Afterwards, we obtain the total number of produced pairs per radiating electron in the D-model by weighting with the bremsstrahlung spectrum, according to
\begin{align}
N_D =  \int_0^1 df \mathscr{P}(\omega') I_\gamma(f,\ell)\,. \label{ricondaN}
\end{align}
For our numerical calculations we cut off the integration domain in Eq.~\eqref{ricondaP} at a sufficiently large upper boundary $b_{\rm max}=3 w_0$, that corresponds to a cross sectional area of $A_{\rm max} = \pi(3 w_0)^2$. Then we distribute the associated portion of $A_{\rm max} / A_\gamma$ bremsstrahlung $\gamma$-photons uniformly over the area $A_{\rm max}$ and calculate the contribution of each trajectory to the pair yield. It is worth mentioning that the maximal value of $\kappa$ [see Eq.~\eqref{kappa}] is experienced by those $\gamma$-photons which pass through the center of the laser focus at $b\approx 0$ and have an energy close to $E_0$.

\subsection{Extension of the D-model:\\ Second generation of created pairs}
\label{TheorySubsectionD}
Our description of the production process in Secs.~\ref{TheorySubsectionB} and \ref{TheorySubsectionC} solely accounts for an initial pair creation event, i.e.~for the first generation of created pairs. Since the particles are generated with high energies and, directly after their creation, are still subject to the strong laser field, they can in principle induce further pair production events. In particular, a created high-energy electron (or positron) counterpropagating a high-intensity laser pulse can emit a $\gamma$-photon through Compton scattering, which afterwards induces another Breit-Wheeler pair creation event. Such pair production showers (or cascades) have been studied intensively in recent years \cite{Bell-Kirk, Elkina, Mironov, Blackburn2017, Pouyez2024}.

By a suitable amendment of our treatment, we shall account---in an approximate way---for the possibility of subsequent pair creation events which are induced by the first generation of pairs. Previous calculations of nonlinear Breit-Wheeler pair production by a monoenergetic $\gamma$-beam colliding with an intense laser pulse \cite{Riconda} have shown that the second generation of pairs starts to become relevant for values $\kappa\gtrsim 4$--5 of the quantum nonlinearity parameter. A subsequent study \cite{Pouyez2024} has revealed that even higher generations of pairs would contribute for $\kappa\gtrsim 16$--20. The latter study assumed linearly polarized, plane-wave laser pulses with sin$^2$-shaped intensity profile. The yield of created pairs was obtained by numerically solving a coupled system of integro-differential equations for the distributions functions of electrons, positrons and $\gamma$-photons (see Sec.~II.C in \cite{Pouyez2024}). The corresponding numerical code has been made available \cite{shower-code}. This way, the numbers $N_1$ and $N_2$ of pairs created in the first and second generation, respectively, were computed, from which enhancement factors $(N_1+N_2)/N_1$ can be obtained. They quantify the amplification of the pair yields due to the second generation over the sole contribution of the first generation.

We have combined the framework developed in \cite{Pouyez2024} with our treatment to estimate the additional contribution to the pair yield when the second generation is included. In a setup with incident electron energy $E_0$ and laser intensity parameter $\xi$, we determine the enhancement factors $(N_1+N_2)/N_1$ for each value of $f$ and $b$ by running the code \cite{shower-code} multiple times. Calculating the enhancement for fixed $\gamma$-photon energy $\omega'=fE_0$ and impact parameter $b$ (this way fixing the maximum field strength $\mathcal{E}_0(b)=\mathcal{E}_0{\rm e}^{-(b/w_0)^2}$ experienced by the $\gamma$-photon) is necessary since the analysis in \cite{Pouyez2024} assumes a monoenergetic $\gamma$-beam and unfocused laser pulse. Accordingly, by running the code separately for each pair of parameter values $\omega'$ and $\xi(b)=|e|\mathcal{E}_0(b)/(m\omega)$, we scan over the bremsstrahlung spectrum and the transverse laser profile. The resulting enhancement factors are inserted into the two-dimensional integral over $f$ and $b$ [see Eqs.~\eqref{ricondaP} and \eqref{ricondaN}], in order to obtain the pair yield including the contribution from the second generation.

Our corresponding results will be shown in Sec.~\ref{ResultsSubsection2} where pair creation in collisions of bremsstrahlung with laser pulses at $\kappa\gg 1$ is considered.

\section{Results and Discussion}\label{Results}

Based on the S- and D-models described above, our goal in this section is to analyze the physical properties of the strong-field Breit-Wheeler pair production in the overcritical regime with $\kappa > 1$. For our numerical studies, a Ti:Sapphire laser with frequency $\omega=1.55$ eV is assumed throughout with intensity parameter $\xi\sim \mathcal{O}(100)$ and pulse duration $\tau\sim\mathcal{O}(10)$\,fs. Unless specified otherwise, the focal beam waist is chosen as $w_0=2 \ \mu$m, corresponding to a Rayleigh length $z_R \approx 15.7 \ \mu$m. The laser pulse collides with bremsstrahlung $\gamma$-photons generated by a few-GeV electron beam of 10 pC total charge. It is assumed that 1\% of the incident electrons emit bremsstrahlung, which is a reasonable amount for the chosen target thickness.

\subsection{Comparison between S-model and D-model at moderately high field strengths}
\label{ResultsSubsection1}

We begin with presenting the results of numerical calculations performed with the S-model of Sec.~\ref{TheorySubsectionB} and the D-model of Sec.~\ref{TheorySubsectionC}. Our aim is to compare the obtained pair production yields at moderately high field strengths $\xi\lesssim 100$ and to characterize the borderline of parameters starting from which the D-model needs to be applied because the attenuation of the decaying $\gamma$-beam cannot be disregarded anymore.

\begin{figure}[ht]
\includegraphics[width=0.5\textwidth]{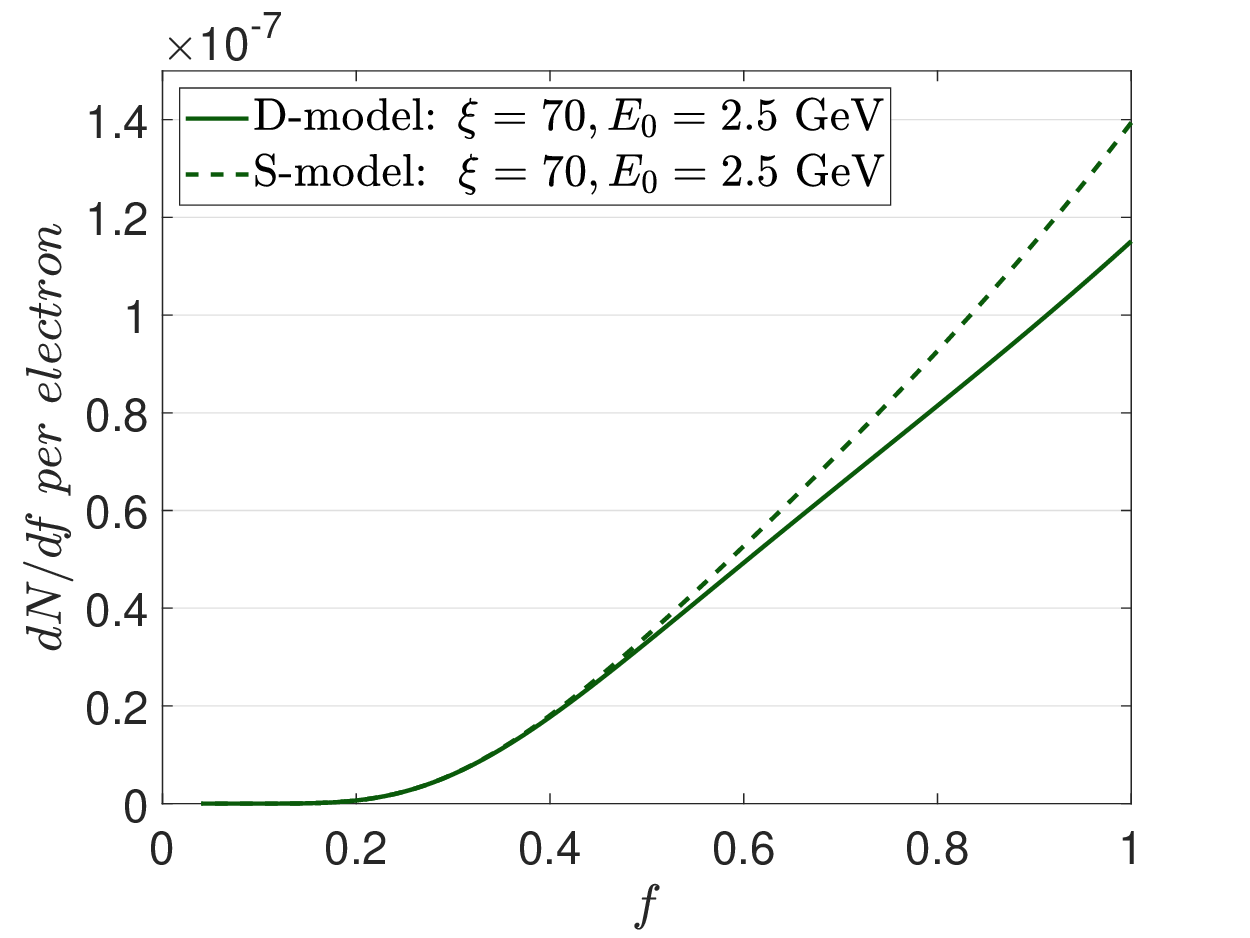}
\caption{\label{figXI70Vergleich} Comparison between the D-model (solid) and S-model (dashed) results for the differential number of created pairs from different bremsstrahlungs photon energies at $\xi = 70$ and $E_0 = 2.5$ GeV.}
\end{figure}

Figure~\ref{figXI70Vergleich} shows the differential number of created pairs (per radiating electron) in dependence on the normalized bremsstrahlung-photon energy. The set of parameters of a planned experiment on strong-field Breit-Wheeler pair production \cite{CALA, Golub2022} is considered here: $E_0=2.5$~GeV, $\xi=70$ (corresponding to an intensity of $I \approx 10^{22}$~W/cm$^2$), and $\tau = 30$ fs. The highest value of the quantum nonlinearity parameter resulting in this scenario is $\kappa\approx 2$ for $f\approx 1$. As the figure displays, the contribution to the pair production yield grows monotoneously with increasing $f$. Up to $f\approx 0.5$ (corresponding to $\kappa\approx 1$), the predictions of the S- and D-models closely agree with each other, but start to slightly deviate afterwards. At the bremsstrahlung endpoint $f\approx 1$, the S-model over\-estimates the pair yields by about 20\%.\footnote{We recall that, at $f=1$, the differential number $dN/df$ would actually sharply drop to zero. However, this feature of a real bremsstrahlung spectrum is not covered by the analytical thin-target formula~\eqref{thin}, as was explained in Sec.~\ref{TheorySubsectionA}.}

When the differential pair numbers are integrated over $f$ and multiplied by the number of radiating electrons, we obtain a total yield of $\approx 0.025$ pairs per laser shot from the D-model and $\approx 0.028$ pairs per laser shot from the S-model. Thus, the $\gamma$-beam decay has only a minor impact on the total yield to be expected in the experiment \cite{CALA}. Accordingly, the obtained numbers from both the S- and D-model are in good agreement with the pair yield of 0.03 reported in \cite{Golub2022}. For later use we note that the S- and D-model predictions for the total yield differ by about 10\% in the present example, which is much less than the difference of about 20\% arising in the spectrally resolved curves in Fig.~\ref{figXI70Vergleich} at the endpoint $f\approx 1$. The reason is that to the total yield also lower $f$-values give significant contributions, where S- and D-model lie (much) closer to each other.

When the quantum nonlinearity of the process is enhanced by increasing the incident electron energy, the differences between S- and D-model grow. Using the D-model for $E_0 = 5$ GeV results in $\approx 0.12$ pairs per laser shot (where the S-model would have given a result of $\approx 0.16$) and for $E_0 = 10$ GeV results in $\approx 0.31$ pairs per laser shot (where the S-model would have given a result of $\approx 0.47$). Thus, for $E_0 = 5$ GeV, corresponding to a maximal value of $\kappa \approx 4$, the S-model overestimates the pair yield already by about $25 \%$, whereas for $E_0 = 10$~GeV (where $\kappa\lesssim 8$) the yield is overestimated by about $50\%$. These numbers indicate that the D-model should be utilized for these parameters to obtain quantitatively reliable predictions.

Apart from the value of the quantum nonlinearity parameter $\kappa$, we emphasize that also the duration of the interaction influences the relation between S- and D-model predictions. Let us demonstrate this by a simple estimate. In the tunneling-like regime of strong-field Breit-Wheeler pair production at $\kappa\ll 1$, the process rate is given by an analytical expression of the form $R \sim \frac{\alpha m^2 \kappa}{\omega'}\,{\rm e}^{-8/(3\kappa)}$. This compact formula still holds approximately for $\kappa\sim 1$ (see, e.g., Fig.5 in \cite{Golub2022}). Taking $\kappa=2$ and $\omega'=2.5$ GeV, the rate becomes $R\sim 10^{-7}m$. As a consequence, the time-integrated rate remains smaller than unity for laser pulse durations $\tau\lesssim 10$ fs. Therefore, the S-model is applicable for few-cycle optical laser pulses with duration of the order or below $\approx 10$~fs in this case.

Another interesting aspect in this context is the scaling of the total pair yield with the laser pulse duration. Reducing the previous value from $\tau=30$ fs to $\tau=20$~fs, the D-model (S-model) predicts $\approx 0.09$ ($\approx 0.10$) pairs per laser shot for $E_0=5$ GeV and $\approx 0.23$ ($\approx 0.32$) pairs per laser shot for $E_0=10$ GeV. These results illustrate (via comparison with the corresponding numbers for 30~fs given above) that, to a good approximation, S-model predictions are proportional to the laser pulse duration. However, in a parameter regime, where the use of the D-model is required, the scaling with $\tau$ becomes more complex and nonlinear. This distinctive feature is a direct consequence of the different mathematical forms of Eqs.~\eqref{N} and \eqref{RicondaSinglePhotonProbability}. 

\begin{figure}[h]
\includegraphics[width=0.5\textwidth]{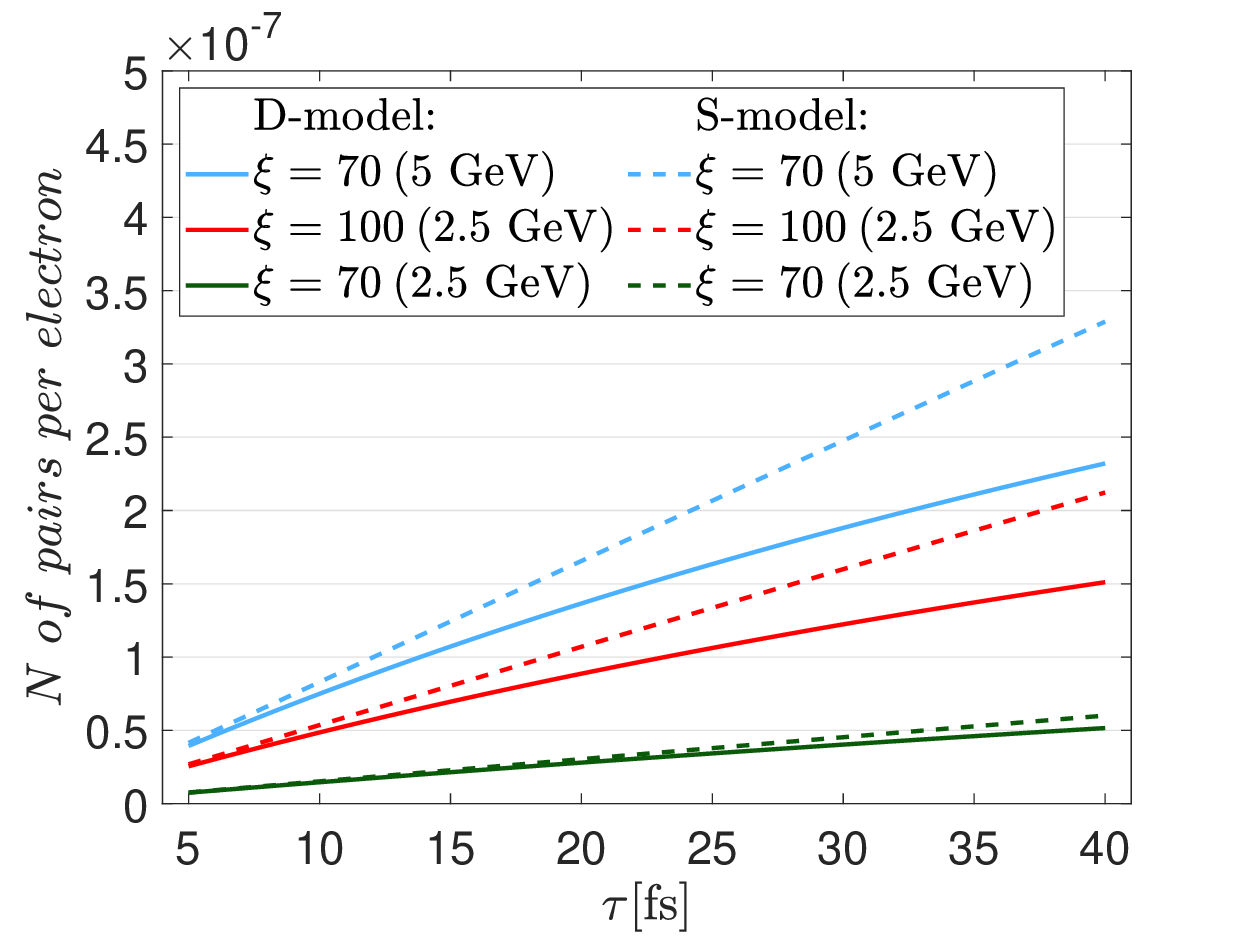}
\caption{\label{tauplot} Dependence of the number of created pairs on the laser duration predicted by the S-model (dashed lines) and the D-model (solid lines) for incident electron energy $E_0 = 5$~GeV at $\xi = 70$ (blue curves), $E_0 = 2.5$ GeV at $\xi = 100$ (red curves) and $E_0 = 2.5$ GeV at $\xi = 70$ (green curves).}
\end{figure}

The relevance of and scaling with the interaction time is illustrated in Fig.~\ref{tauplot}. It depicts the number of created pairs in dependence on the laser pulse duration for $E_0 = 2.5$ GeV at $\xi = 70$, $E_0 = 2.5$ GeV at $\xi = 100$, and $E_0 = 5$ GeV at $\xi = 70$. For the S-model, we observe a linear dependency on the laser duration for both energies, whereas in the D-model, photon decay while penetrating the laser beam leads to a sublinear growth in the number of pairs with increasing laser duration. For very short pulse durations, the predictions from both models coincide in all three cases. While for $E_0=2.5$ GeV and $\xi=70$, the difference between S- and D-model remains moderate throughout the considered $\tau$-range, the discrepancy grows stronger for larger values of $E_0$ and $\xi$. For instance, at $E_0 = 2.5$ GeV and $\xi = 100$ the S-model overestimates the total yield for $\tau = 20$ fs by about 20\%, for $\tau = 30$ fs already by about 30\%, and for $\tau = 40$~fs quite sizeably by about 40\%. Nearly the same percentage values of relative overestimation by the S-model over the D-model result for the parameter combination $E_0 = 5$~GeV and $\xi = 70$.

\subsection{Pair production in the overcritical field regime}
\label{ResultsSubsection2}

In this subsection, we utilize our D-model and apply Eq.~\eqref{ricondaN} to obtain several important insights into the Breit-Wheeler pair creation process in the overcritical regime with  $1<\kappa\lesssim 30$. Also the contributions from the second generation of created pairs will be considered. In contrast to the previous Sec.~\ref{ResultsSubsection1}, here the attention is put on high laser intensities above $10^{22} \ \mathrm{W/cm^2}$. These values are achievable in the foreseeable future at the laser facilities such as ELI-NP in Romania, where the investigation of the Breit-Wheeler pair production is firmly embedded into the planning \cite{ELI}. In the following, we shall therefore use a laser pulse duration of $\tau=20$ fs, as a typical value envisaged at this facility.

\subsubsection{Spectrally resolved pair yields}

\begin{figure}[b]
\includegraphics[width=0.5\textwidth]{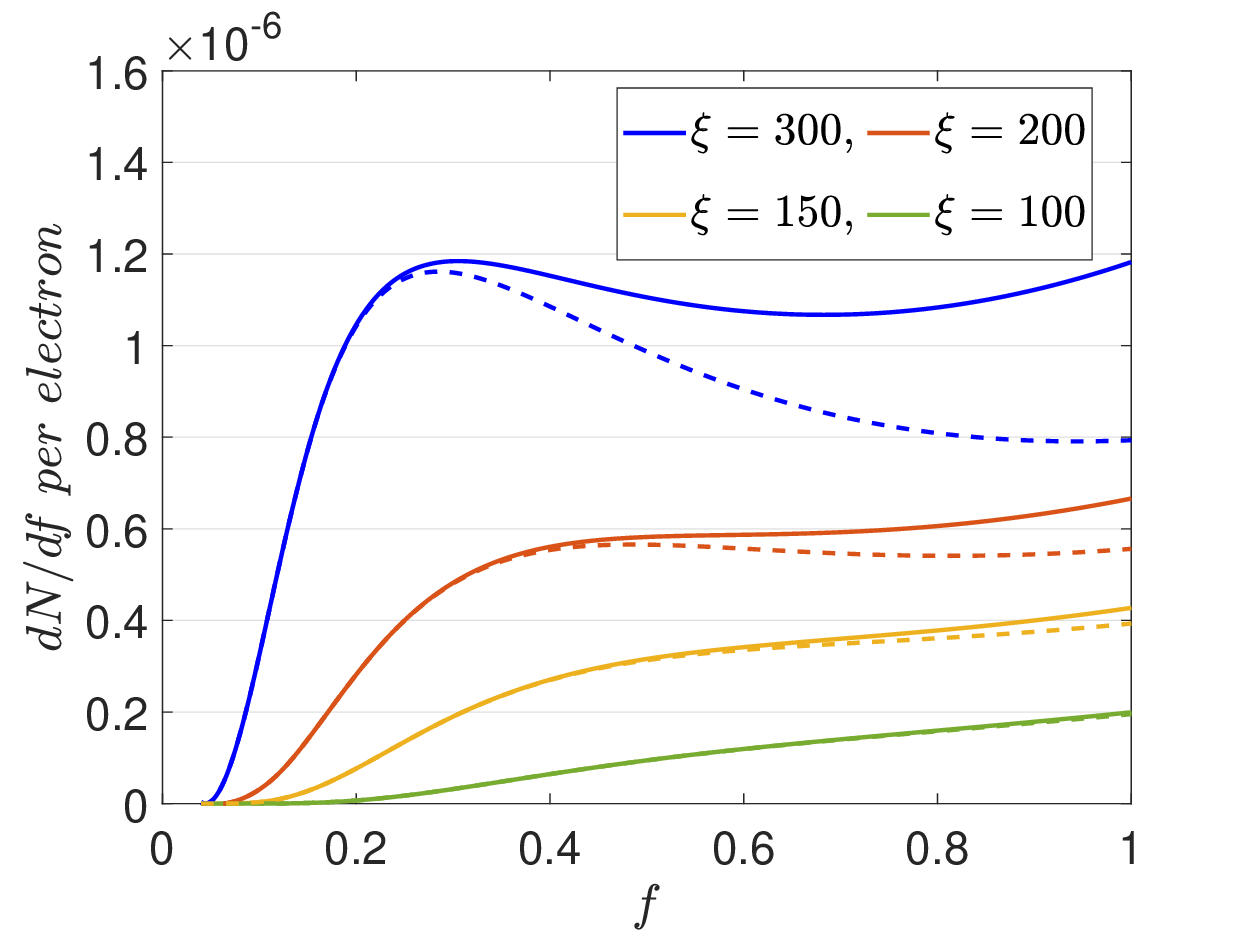}
\caption{\label{figZusatzplot} Differential number of created pairs from different bremsstrahlung photon energies at incident electron energy $E_0 = 2.5$ GeV for different values (from bottom to top) of the laser intensity parameter $\xi \in \{100,150,200,300\}$ as indicated. The dashed lines show the D-model results for the first generation of pairs; the solid lines take the additional contributions from the second pair generation into account.}
\end{figure}

By utilizing the D-model we can analyse the contribution to the number of created pairs from different $\gamma$-photon energies within the bremsstrahlung spectrum. These contributions are depicted in Fig.~\ref{figZusatzplot} for different values of $\xi$ between 100 and 300 at a fixed incident electron energy of $E_0 = 2.5$ GeV. The dashed lines refer to the first pair generation, whereas the solid lines show the sum of the first and second pair generations.

We first discuss the properties of the curves for the first pair generation. As we go to high values of $\xi$ we see a clear contrast in the distributions compared to the case in Fig.~\ref{figXI70Vergleich}. For $\xi = 100$, we observe as before a strictly monotonic increase in the number of created pairs as $f$ grows. In contrast, for higher values of $\xi$, a maximum emerges, indicating a peak contribution at specific $\gamma$-photon energies. For $\xi = 150$, the curve already shows a slight bend, for $\xi = 200$, we observe a flat global maximum at $f \approx 0.48$, which develops into a pronounced maximum at $f \approx 0.28$ for $\xi = 300$. The formation of those maxima results from three effects which influence the creation of first-generation pairs. Firstly, for the parameters under consideration, the local rate of pair creation still approaches to an exponential dependence and thus increases steeply with $\kappa$, which is proportional to the product of $\xi$ and $f$. Thus, for $\xi \lesssim 100$, the exponential rate damping is still strong and therefore the number of pairs grows as $f$ increases, whereas for high values of $\xi$, the field dependence of the rate becomes much flatter (see also Fig.~\ref{NofPairsPlot} below). Secondly, the number of produced bremsstrahlung photons is much higher for low values of $f$, as depicted in Fig.~\ref{fig0}, which starts playing a crucial role in the pair production when the rate growth has left the very steep exponential region and entered into the flatter portion. Lastly, the exponential decay of the bremsstrahlung photons in the D-model is much more pronounced for high values of $\kappa$, which also favors low values of $f$ for the creation of pairs. It is therefore, that we observe the emergence of a maximum at lower values of $f$, because the increase of the rate with higher values of $f$ is counteracted and eventually overcompensated by the falling distribution of bremsstrahlung photons and their decay for high values of $\xi$.

Next we consider the additional contribution from the second generation of pairs. As the solid curves in Fig.~\ref{figZusatzplot} show, these contributions increase when $f$ grows. While for $\xi=100$ the enhancement of the pair yield due to the second generation is still marginal, for $\xi=150$, 200 and 300 the enhancement factors at $f\approx 1$ amount to 1.08, 1.19 and 1.49, respectively. As a result of the enhanced pair yields, the solid curve for $\xi=200$ does not feature a maximum but continues to grow slowly. Instead, a pronounced maximum at $f\approx 0.32$ remains in the solid curve for $\xi=300$, which is followed by a shallow minimum at $f\approx 0.68$ and a subsequent distinct increase. Furthermore, from the results for $\xi=200$ and 300 we can infer that the plain D-model of Eq.~\eqref{ricondaP} provides a good approximation to the differential pair yield $dN/df$ up to quantum nonlinearity parameters of $\kappa\approx 6$, where the additional contributions from the second pair generation remain below $\approx 20$\%.

\begin{figure}[t]
\includegraphics[width=0.5\textwidth]{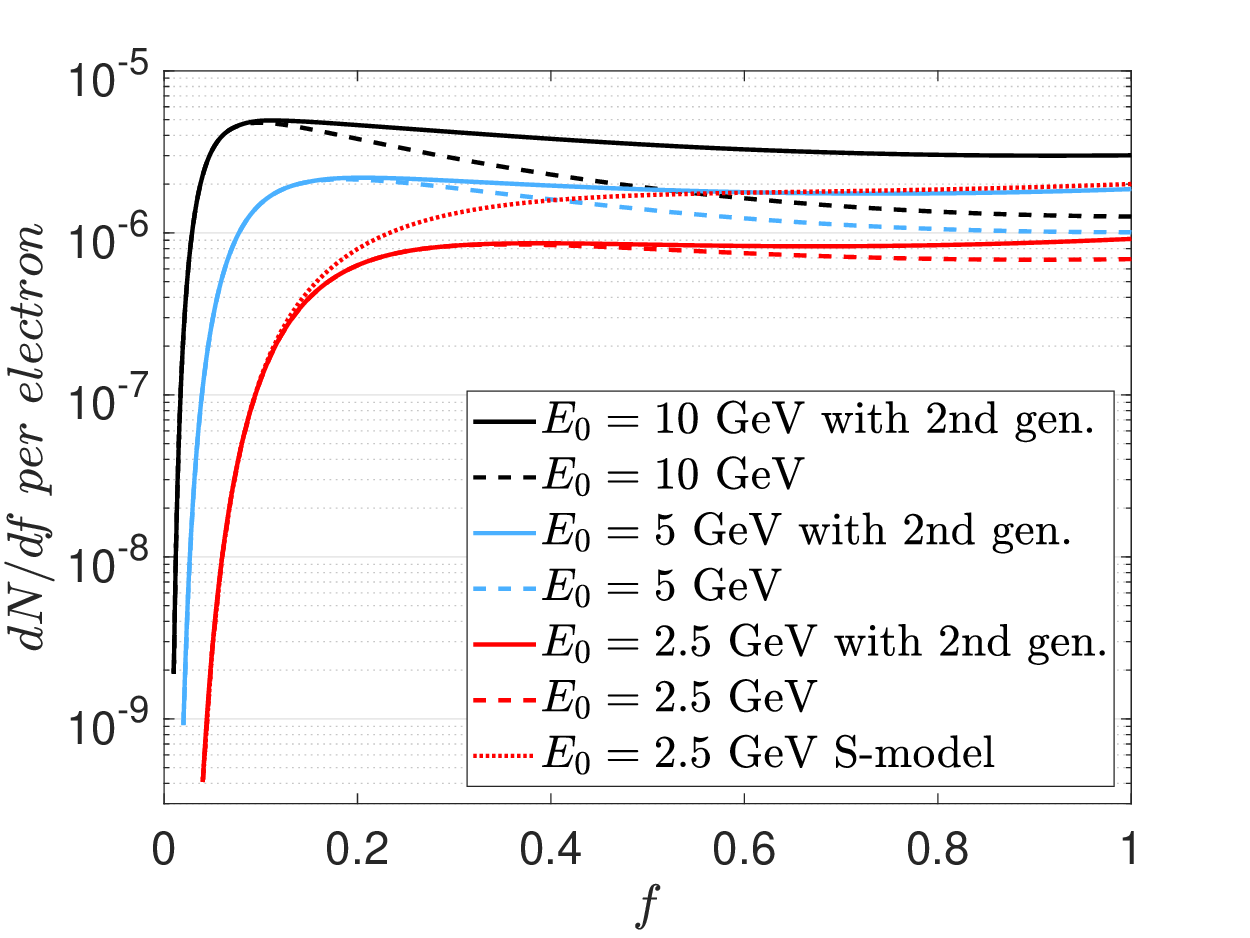}
\caption{\label{xi250plot} Contributions to the number of created pairs from different  bremsstrahlung photon energies at $\xi=250$. The red, blue and black curves (in this order) represent different values of the incident electron energy $E_0 \in \{2.5, 5, 10\}$~GeV. The dashed lines show the results for the first generation of pairs obtained from the D-model; the solid lines also include the second generation of pairs. For comparison, the dotted line represents the result for the first pair generation at $E_0 = 2.5$~GeV predicted by the S-model.}
\end{figure}

A similar observation can be made when we compare the impact of the energy of bremsstrahlung photons for different values of the incident electron energy. In Fig.~\ref{xi250plot} the contribution to the number of created pairs is plotted for $E_0 = 2.5$ GeV (red), $E_0 = 5$ GeV (blue) and $E_0 = 10$ GeV (black) at fixed $\xi = 250$. Maxima in the number of created first-generation pairs (dashed curves) arise at $f \approx 0.35$ for $E_0 = 2.5$ GeV, $f \approx 0.185$ for $E_0 = 5$ GeV, and most distinctly for $E_0 = 10$ GeV at $f \approx 0.0975$. When the second pair generation is taken into account (solid curves), these maxima remain but become less pronounced. Here, the enhancement factors at $f\approx 1$ are 1.33, 1.85 and 2.39 for $E_0=2.5$~GeV, 5~GeV and 10~GeV, respectively. As before, the relative contributions from the second pair generation lie below $\approx 20$\% up to $\kappa\approx 6$.

For comparison, the S-model prediction for the first pair generation at $E_0 = 2.5$~GeV is represented in Fig.~\ref{xi250plot} by the dotted red line. It agrees with the D-model result up to $f\approx 0.15$, corresponding to $\kappa \approx 1.1$, but sizably overestimates the yield for higher bremsstrahlung photon energies. In particular, no maximum emerges along the S-model curve. The S-model results for the higher electron energies are not shown in Fig.~\ref{xi250plot} because they would deviate even more strongly from the D-model. We note that these curves would feature maxima at larger $f$ values because of the counteracting trends between the increase of the rate and the decrease of the bremsstrahlung spectrum. An additional effect contained in the D-model description is the attenuation of the $\gamma$-beam. In result, the maxima of the blue and black dashed lines are located at smaller $f$ values than one would expect based on an S-model consideration. 

\begin{figure}[t]
\includegraphics[width=0.5\textwidth]{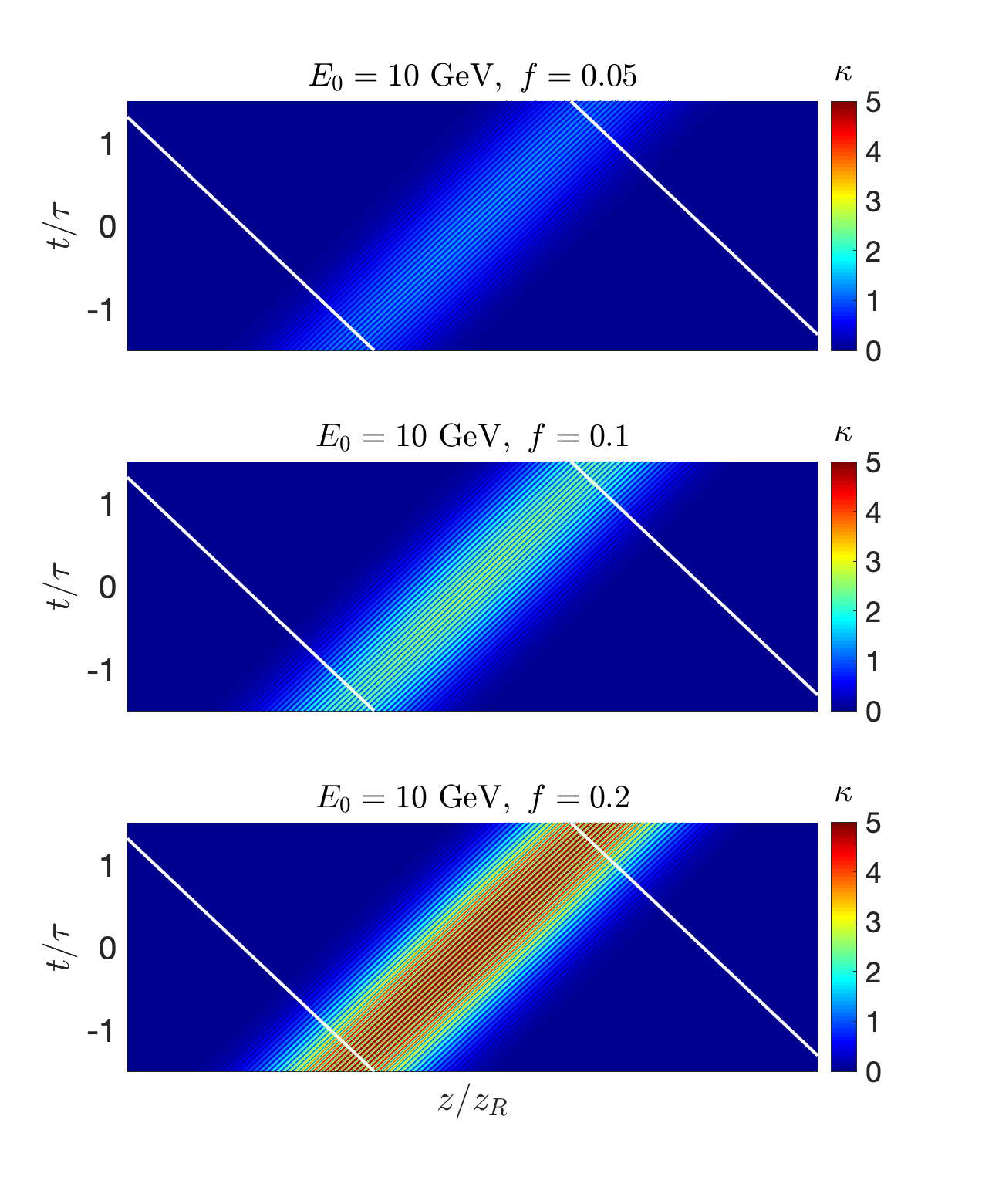}
\vspace{-1cm}
\caption{\label{AlinaKappaPlot} Time and space dependent quantum nonlinearity parameter $\kappa(t,0,z,f)$ at the point $\rho=0$ for photons emitted by $10$ GeV electrons at $f=0.05$ (upper panel), $f=0.1$ (middle panel) and $f=0.2$ (lower panel). Moreover, $\xi=250$ is chosen and the integration region as given in Eq.~(\ref{NGauss}) is encompassed between white lines.}
\end{figure}

As an illustration, Fig.~\ref{AlinaKappaPlot} depicts the values of the spacetime-dependent quantum nonlinearity parameter $\kappa(t,\rho, z,f)$ at $\rho=0$ for 10 GeV incident electrons and the integration region given in Eq.~(\ref{NGauss}). In the upper panel the local values of $\kappa$ for very small photon energy with $f=0.05$ is exposed. Here, the number of created particles is expected to be low as $\kappa$ is of the order of one. On the other hand, for $f=0.1$ the values presented in the middle panel overpass the critical value of $\kappa \approx 1$ and the expected pair yield per $\gamma$-photon is significantly increased, as the heavy exponential damping characteristic for the tunneling-like regime at $\kappa \ll 1$ is largely softened here, so that the course of the pair production rate is distinctly flattened. This observation combined with the high number of produced bremsstrahlung photons (see the graph in Fig.~\ref{fig0}) is responsible for the emergence of the maximum in Fig.~\ref{xi250plot}. Lastly, the quantum nonlinearity parameter for $f=0.2$ is shown in the lower panel of Fig.~\ref{AlinaKappaPlot}. Here, $\kappa \approx 6$ is reached leading to a high rate of the particle creation. However, as the number of bremsstrahlung $\gamma$'s is much lower than for $f=0.1$ (see Fig.~\ref{fig0}) and their attenuation in the process is stronger, the resulting outcome for particle production is reduced.

\subsubsection{Total pair yields}

\begin{figure}[t]
\includegraphics[width=0.5\textwidth]{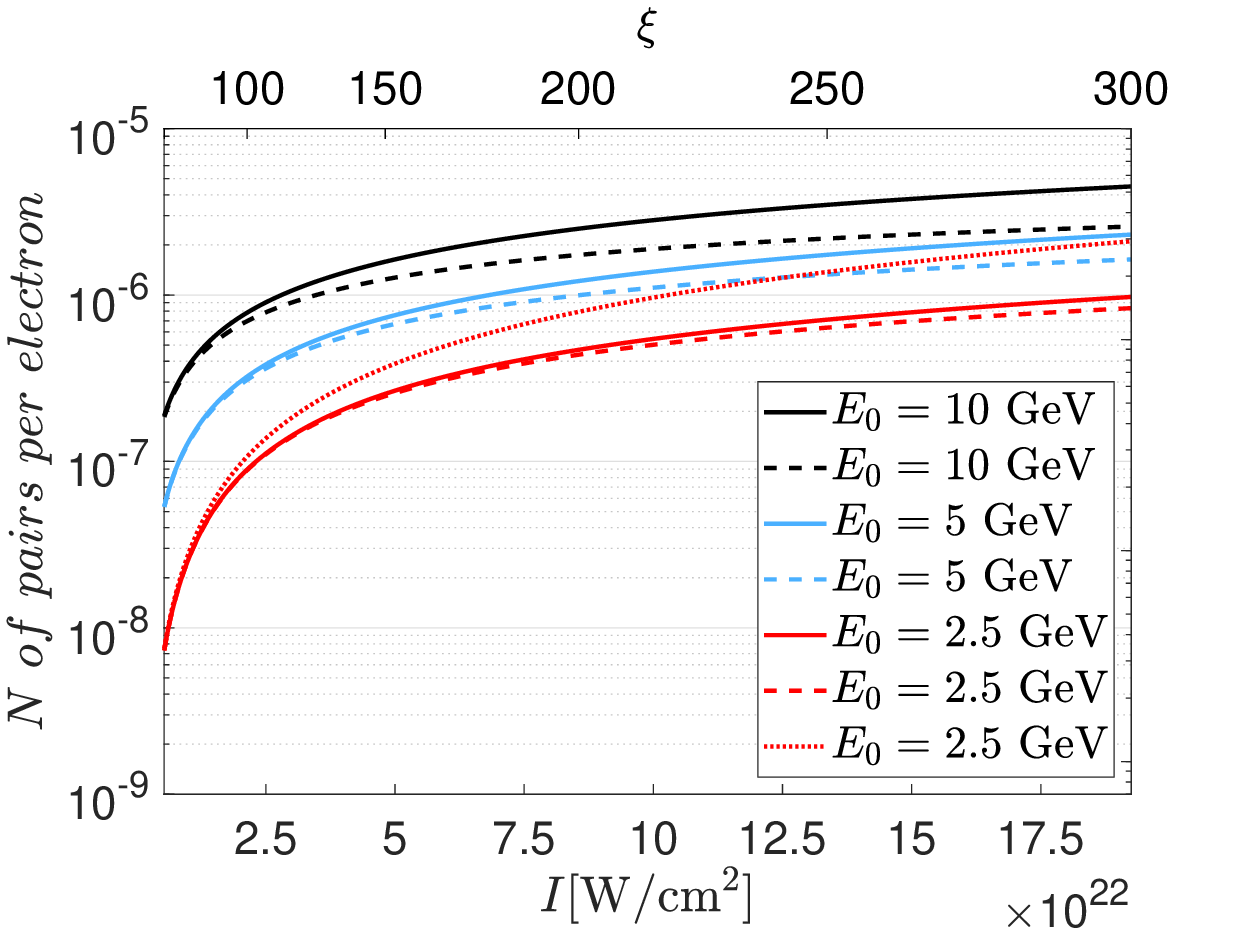}
\caption{\label{NofPairsPlot} Dependence of the number of created pairs on the laser intensity for various incident electron energies, color-coded with red, blue and black lines for $E_0 = 2.5$, 5, and 10~GeV, respectively. The dashed lines show the contribution from the first generation of pairs using the D-model; the solid lines additionally include the second generation of pairs. For comparison, the dotted line represents the S-model prediction for the first pair generation at $E_0 = 2.5$ GeV.}
\end{figure}

In Fig.~\ref{NofPairsPlot} the total number of pairs produced by a single radiating electron is considered in the intensity range $\xi \in [50,300]$. Moreover, three different values of the incident electron energy $E_0 \in \{2.5, 5, 10\}$ GeV are studied and depicted as red, blue and black dashed (solid) lines for the first (first + second) pair generation, correspondingly. It can be seen that the increase in the intensity for the lowest electron energy provides an improvement in the number of created pairs by two orders of magnitude. A pronounced rate growth is also visible, however to a smaller extent, for $E_0 = 5$ GeV and $E_0 = 10$~GeV. The steepest increase occurs up to $\xi\approx 150$, from where on the rate growth becomes significantly flatter.

The S-model prediction for the first pair generation would lead to a strong overestimation of the rate practically throughout the whole range of considered laser intensities. This is illustrated by the red dotted line for $E_0=2.5$ GeV. It approximately agrees with the D-model rate up to $\xi\approx 70$, which is associated with a maximal $\kappa \approx 2.1$, and strongly deviates for the higher intensities. The deviations to the D-model prediction would be even larger for the higher incident electron energies, which are therefore not included in the figure.

At this point we emphasize that the $\kappa$ values, for which the S- and D-models coincide, are lower for the spectrally resolved pair yields (see discussion of Figs.~\ref{figXI70Vergleich} and \ref{xi250plot}) than for the total pair yields. As already mentioned before this is because one integrates over the spectrally resolved curve in order to get the total pair yield, and thus includes also those parts of the bremsstrahlung spectrum where both models are in good agreement.

The total pair yield (including the second generation) in an envisaged experiment can be obtained by taking the number of interacting $\gamma$-photons into account. As mentioned above, we assume that the incident electrons come in bunches of 10 pC total charge and that $1\%$ of them will emit bremsstrahlung. Thus in the range $\xi \in [50,300]$, for $E_0=2.5$ GeV, from 0.005 to 0.61 pairs per shot can be expected, depending on the applied intensity. Next, by increasing the incident electron energy to 5 GeV \cite{LWFA}, the respective pair yields can be amplified significantly to 0.03 and to 1.44 pairs per shot. Lastly, the amount of 0.12 (at $\xi=50$) and up to 2.81 (at $\xi=300$) created pairs is achievable in the most optimistic scenario with $E_0=10$ GeV. For comparison, the yields amount to 0.52, 1.02 and 1.61 first-generation pairs per shot at $\xi=300$ and $E_0=2.5$~GeV, 5~GeV and 10~GeV, respectively.

Given the enhancement due to the second pair generation for large values of $E_0$ and $\xi$, the question arises whether also contributions from the third pair generation are relevant. It has been shown in \cite{Pouyez2024} that inclusion of the second generation is sufficient for quantum nonlinearity parameters up to $\kappa\approx 16$--20. We have checked that, in our case, the contribution to the first generation from values of $\kappa\ge 16$ is about 8\% for the largest parameter combination of $E_0=10$~GeV and $\xi=300$. Hence we may conclude that, in the parameter range under consideration, the third pair generation would give only minor additional contributions to the total pair yields.

Continuing our analysis of Fig.~\ref{NofPairsPlot} by comparing the yield between 2.5 GeV and 5 GeV, we observe an almost fourfold raise in the number of created (first + second generation) particles at $\xi = 100$ from 0.056 pairs to 0.20 pairs per shot. At $\xi = 250$, doubling the energy from 2.5 to 5 GeV still results in increasing the yield from 0.44 pairs to 1.09 pairs per shot. For 10 GeV we observe 0.49 pairs at $\xi = 100$ and up to 2.18 created pairs for $\xi = 250$. Hence, the numbers presented here drastically outperform the predictions for 'discovery experiments' planned in the nearest future (see Sec.~\ref{ResultsSubsection1}) and would thus allow for precision tests of the underlying theory.

The plain D-model of Eq.~\eqref{ricondaN} provides a good approximation to the total pair yields in Fig.~\ref{NofPairsPlot} up to maximal quantum nonlinearity parameters of $\kappa\approx 12$--15, where the relative additional contributions from the second pair generation remain below $\approx 20$\%. These $\kappa$-values are distinctly larger than those found in the discussion of the spectrally resolved pair yields $dN/df$ in Figs.~\ref{figZusatzplot} and \ref{xi250plot}. Similarly as before when comparing the S- and D-model predictions, this is because the total pair yield involves an integration over the whole bremsstrahlung spectrum, which contains $f$-regions where the second pair generation does not give sizeable contributions.

\begin{figure}[t]
\includegraphics[width=0.5\textwidth]{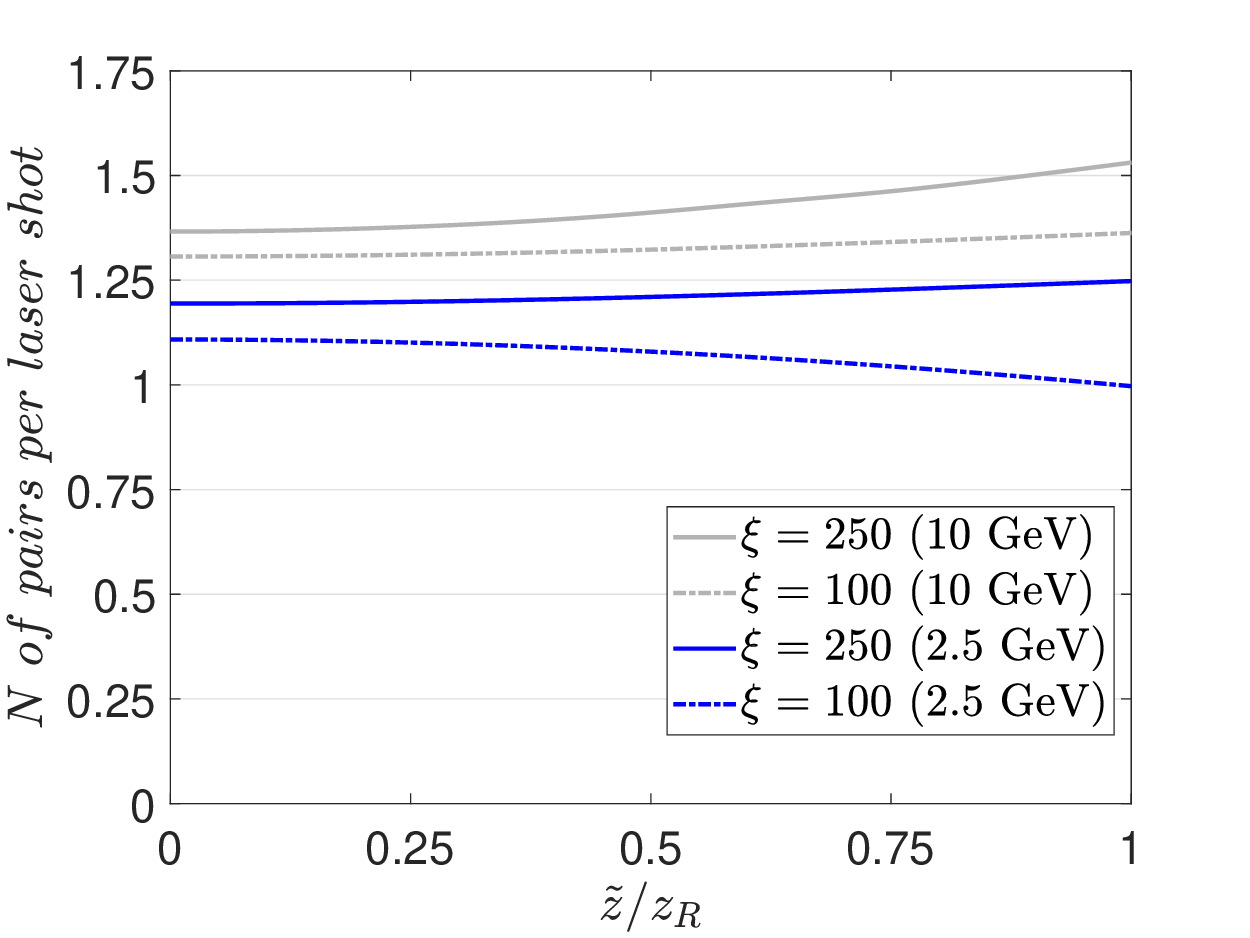}
\caption{\label{offsetPlot} Number of created pairs per laser shot as a function of the relative offset of the $\gamma$-photon trajectories for different values of $\xi \in \{100, 250\}$ at $E_0 =$ 2.5 GeV and 10 GeV. For presentation purposes the lowest curve was scaled by an amplification factor of 20, and the two curves in the middle were amplified by a factor of 3. The considered maximum offset coincides with the upper boundary of our integration region [see e.g. Eq.~\eqref{NGauss2}].}
\end{figure}

At this point, we can quantify the appropriateness of the approximation in Eq.~\eqref{traj} where the longitudinal offset in the $\gamma$-photon trajectory was set to zero, $\tilde{z}=0$. It means that the considered $\gamma$-photons reach the center of the laser focus at time $t=0$, where the laser field strength is maximum. When the trajectory is taken more generally as $\vec{r}(t)= - t \hat{e}_z + \tilde{z} + b \hat{e}_\rho$, those $\gamma$-photons with $\tilde{z}>0$ ($\tilde{z}<0$) hit the laser pulse maximum by $\tilde{t}=\tilde{z}/2$ earlier (later). An average over the relevant range of $\tilde{z}$-values would need to be taken in a more accurate treatment to calculate the pair yield from a bremsstrahlung $\gamma$-burst. The dependence of the total number of pairs per laser shot on the longitudinal offset is depicted in Fig.~\ref{offsetPlot} for $0 \le \tilde{z} \le z_R$. Here, the first generation of produced pairs is considered. One can see that the curves are very flat, implying that it is not strictly necessary to take an average; accordingly, the simplifying assumption $\tilde{z}=0$ in Eq.~\eqref{traj} works well in the considered range of parameters. The largest relative differences arise for $E_0 = 2.5$ GeV and $\xi = 100$, where the pair yield is  reduced by about 2.5\% (10\%) when $\tilde{z}=z_R/2$ ($\tilde{z}=z_R$); for $E_0 = 10$ GeV and $\xi = 250$ the pair yield is instead increased by about 3\% (12\%) for the same values of $\tilde{z}$. The approximation works well for both large values of $\xi$ as well as large values of $E_0$. 

The reason for the slight decrease or increase of the pair number for larger $\tilde{z}$ is that the laser pulse is then hit by the $\gamma$-ray when its center has passed the focal point. The correspondingly reduced field intensity leads to a decrease of the pair number for $\xi=100$, $E_0=2.5$ GeV where the value of $\kappa$ is still moderate and the behavior of the production rate is exponential-like. Instead, for large $\kappa\gg 1$, the rate $R(\kappa)$ grows only rather slowly, so that the reduction of the field intensity plays a minor role. Here, the enhanced interaction volume, associated with the transversally broadened extension of the laser pulse, causes the dominant effect, leading to an increase of the pair number.

In result, when averaging over the offsets, we see only very slight deviations in the yield compared to the approximated case: For $E_0 = 2.5$ GeV the resulting yield is lower by about 3.4\% for $\xi = 100$ and higher by about only 1.6\% for $\xi = 250$ than its value at $\tilde{z} = 0$. For $E_0 = 10$ GeV the yield increases by about 1.6\% for $\xi = 100$ and by about 4.2\% for $\xi = 250$. From this we see that our simplification gives valid results.

\begin{figure}[t]
\includegraphics[width=0.5\textwidth]{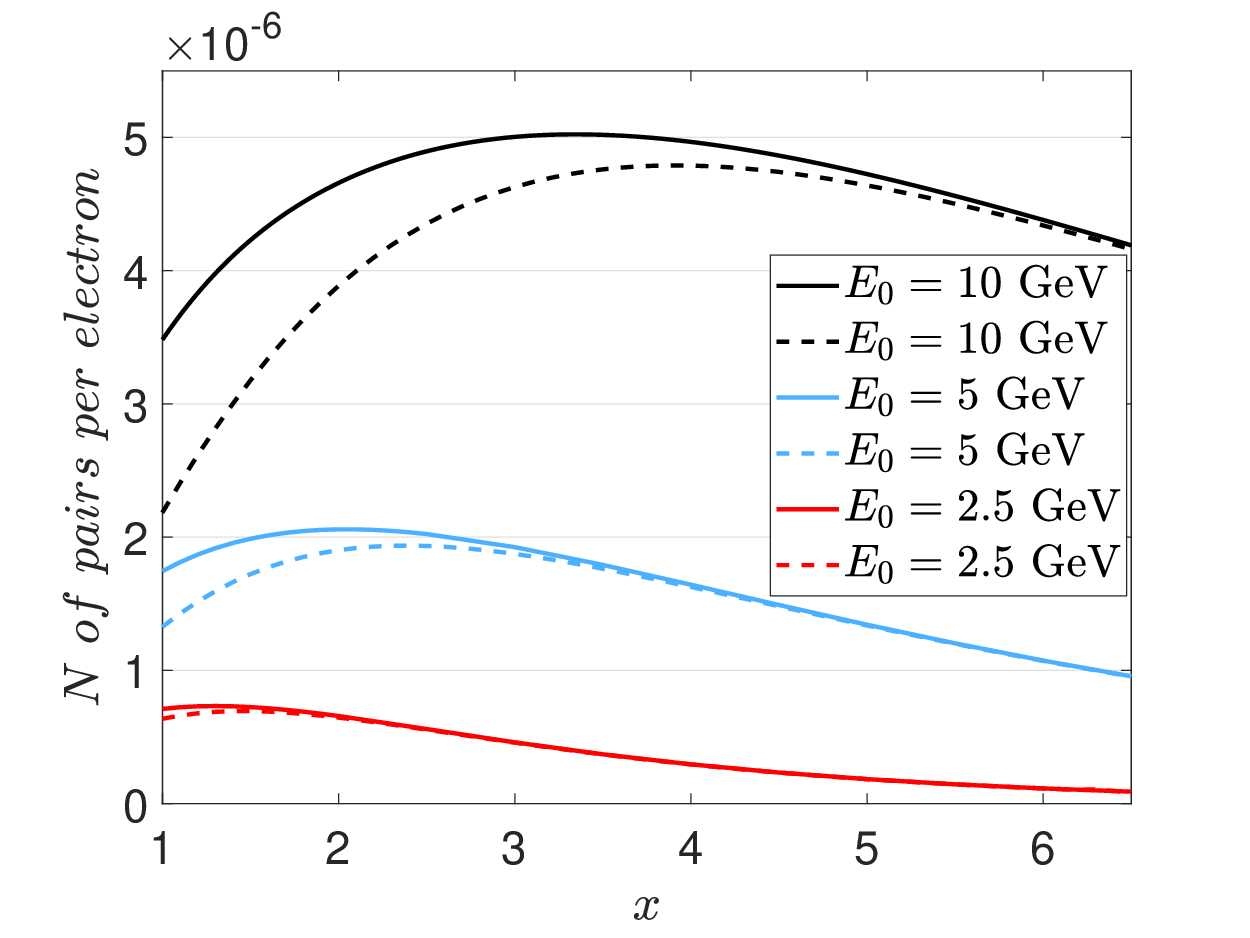}
\caption{\label{xPlot} Number of pairs as a function of the degree of laser focusing obtained by setting $\xi = 250/x$ and $w_0=2x \ \mu$m and keeping the laser energy constant. The red, blue and black colors refer to different values of the incident electron energy $E_0 \in \{2.5, 5, 10\}$ GeV. The dashed lines show the results for the first generation of pairs using the D-model; the solid lines include in addition the second generation of pairs.}
\end{figure}

The influence of tighter or looser laser focusing is displayed in Fig.~\ref{xPlot}. To that end, the energy of the laser pulse
\begin{equation}
E_L\propto \xi^2 w_0^2 = \left(\frac{250}{x} \right)^2 (2x \, \mu  \text{m})^2 =\mathrm{constant}
\end{equation}
is kept constant and the scaling parameter $x$ is varied between 1 and 6.5, leading to higher intensity accompanied by narrower beam waist and visa versa. Accordingly, the scaled intensity parameter $\xi =250/x$ in the range $[39,500]$ with the corresponding beam radius $w_0=2x \ \mu$m in the intevall $[2\,\mu \text{m}, 13\,\mu \text{m}]$ was studied for the incident electron energies $E_0=2.5$ GeV (red), $E_0=5$ GeV (blue) and $E_0=10$ GeV (black). 

Interestingly, the D-model outcomes for the first generation of pairs (dashed lines) show optimal values for the focusing with maximum yield at $x \approx 1.5$ for $E_0=2.5$~GeV, at $x \approx 2.4$ for $E_0=5$~GeV and at $x \approx 3.9$ for $E_0=10$~GeV. On the one hand one might have naivly expected that higher $\xi$ and therefore higher intensities would always lead to higher yields, as depicted in Fig.~\ref{NofPairsPlot} because of the exponential dependency of the rate which rises steeply when $\xi$ grows. But on the other hand the interaction volume decreases for smaller $x$, so that fewer $\gamma$ photons can participate in the creation of pairs; this effect scales quadratically with $x$. For moderate values of $\kappa$, the $x$-dependence of the pair yield is dominated by the depencency of the rate. But for a parameter range where $\kappa$ largely exceeds 1, the rate depency is distinctly flattened and both counteracting effects can compete with each other. Additionally, there is the feature of the D-model that $\gamma$-photons penetrating the laser pulse decay over time which further reduces the normally beneficial effect of higher intensities on the yield (see Fig.~\ref{NofPairsPlot} and Fig.~\ref{xi250plot}). This leads to an even greater necessity for a large interaction volume and correspondingly decreased laser intensity, because then the influence of the photon decay is reduced. As a result, the obtained positions of the focusing optimum in the D-model decrease sublinearly with the increase of $E_0$.

This feature remains when the contributions from the second pair generation are included (solid lines). These contributions increase for higher electron energies, being almost negligible for 2.5 GeV and most prononunced for 10 GeV, and also increase with tighter focusing, since the values of $\kappa$ are higher then. The relative difference between the black dashed and solid lines amounts to $\approx 20$\% at $x\approx 2$, which corresponds to a maximal value of $\kappa\approx 15$ in the bremsstrahlung-laser collision. Although the second generation can enhance the pair yield substantially, it is not large enough to change the overall shape of the curves. Instead the larger interaction volume still plays the crucial role for maximizing the total yield. In result, the inclusion of the second generation only leads for a small shift of the maxima towards a little tighter focusing, with the maximum yield located at $x \approx 1.3$ for 2.5~GeV, $x \approx 2.2$ for 5~GeV and $x \approx 3.3$ for 10~GeV. This implies in summary, that maximum focusing is not recommended, but instead it is adviced to aim for specific focusing depending on the incident electron energy to maximise the number of created pairs.

\section{Conclusions}\label{Conclusions}

Nonlinear Breit-Wheeler pair production in collisions of a focused high-intensity laser pulse with a beam of high-energy $\gamma$-photons from bremsstrahlung has been discussed in the overcritical regime $1<\kappa \lesssim 30$, which will be achievable at the next generation laser facilities with intensities of $\sim 10^{23} \ \mathrm{W/cm^2}$. The presented results complement our previous study \cite{Golub2022} of the intermediate interaction regime around $\kappa \approx 1$. To this end, a theoretical model was used that accounts for the attenuation of the brems\-strahlung beam due to the pair production process. This D-model coincides with our previous approach in the limit of moderate laser intensities and short pulse durations, but also applies to the overcritical regime. In particular, the predictions for the total pair yield from both models are in close agreement up to $\kappa\approx 2$, where the CALA experiment \cite{CALA} is going to operate.

For higher $\kappa$ values the $\gamma$-beam decay has a sizeable effect. One of its consequences is that the first-generation pair yield scales sublinearly with the laser pulse duration. In the considered bremsstrahlung-laser collisions, the D-model was found to represent a good approximation up to $\kappa\approx 6$ for the spectrally resolved pair yields and up to maximal values of $\kappa\approx 12$--15 for the total (bremsstrahlung-integrated) pair yields. However, for larger quantum nonlinearity parameters, the contributions from the second generation of created pairs play a relevant role, which was accounted for within an approximated framework based on Ref.~\cite{Pouyez2024}. While the $\gamma$-beam decay tends to lower the pair production, the additional contributions from the second pair generation have an enhancing effect that grows with increasing $\kappa$.

It was shown that in an experiment combining bremsstrahlung $\gamma$-photons emitted from an incident 10~pC beam of 10 GeV electrons with a 20\,fs, $10^{23}$\,W/cm$^2$ laser pulse, about 1.7 pairs could be generated per shot (to which the second pair generation gives a relative contribution of approximately 30\%). Hence, if the repetion rate of the laser system is 0.1 Hz, about 600 pairs could be produced per hour. This would allow for precision measurements with increased statistical accuracy and deeper understanding of the underlying vacuum structure. 

Regarding the impact of bremsstrahlung photons from different parts of the spectrum it was shown that, in contrast to Ref.~\cite{Golub2022}, also $\gamma$-photons of relatively low energy play an important role in the particle creation. Accordingly, by the considered setup a rather broad range of $\kappa$ values can be probed simultaneously. Finally, the impact of the laser focusing at fixed laser pulse energy was elucidated, with the outcome that in the overcritical regime there is an optimal focal spot size: Beyond this point, focusing even more tightly to reach higher peak intensity would reduce the pair production yield.

\begin{acknowledgements}
This work has been funded by the Deutsche For\-schungsgemeinschaft (DFG) under Grant No.~392856280 within the Research Unit FOR 2783/2. Some of the numerical computations were carried out by using the code~\cite{shower-code} for photon-seeded pair production of M.~Pouyez.
\end{acknowledgements}


\end{document}